\documentclass[12pt, twocolumn]{aastex61}

\usepackage{multirow}
\usepackage{ulem}
\newcommand{\ml}{\bf \color{green}}
\def\rhessi{{\textit{RHESSI}}}
\def\ovsa{{\textit{OVSA}}}
\def\mw{{microwave}}

\def\mwe{{microwave emission}}
\def\gs{{gyrosynchrotron}}

\def\lf{{ _{\rm lf}}}

\received{}
\revised{}
\accepted{\today}
\submitjournal{ApJ}

\shorttitle{Coronal Volume of Energetic Particles from Microwave Imaging}
\shortauthors{Fleishman et al.}

\begin{document}


\title{
The coronal volume of energetic particles in solar flares as revealed by microwave imaging} 

\author{Gregory D. Fleishman$^1$}

\author{Maria A. Loukitcheva$^{1,2,3}$}

\author{Varvara Yu. Kopnina$^{3}$}

\author{Gelu M. Nita$^1$}

\author{Dale E. Gary$^1$}

\altaffiliation{$^1$Physics Department, Center for Solar-Terrestrial Research, New Jersey Institute of Technology
Newark, NJ, 07102-1982}

\altaffiliation{$^2$Saint Petersburg branch of Special Astrophysical Observatory, Pulkovskoye chaussee 65/1,
St. Petersburg 196140, Russia}

\altaffiliation{$^3$Saint Petersburg State University, 7/9 Universitetskaya nab., St.
Petersburg, 199034 Russia}

\begin{abstract}

The spectrum of \gs\ emission from solar flares generally peaks in the \mw\ range.  Its optically-thin, high-frequency component, above the spectral peak, is often used for diagnostics of the nonthermal electrons and the magnetic field in the radio source. Under favorable conditions, its low-frequency counterpart brings additional, complementary information about these parameters as well as thermal plasma diagnostics, either through \gs\ self-absorption, free-free absorption by the thermal plasma, or the suppression of emission through the so-called Razin effect.  However, their effects on the low-frequency spectrum are often masked by spatial nonuniformity. To disentangle the various contributions to low-frequency \gs\ emission, a combination of spectral and imaging data is needed. To this end, we have investigated Owens Valley Solar Array (\ovsa) multi-frequency images for 26 solar bursts observed jointly with Reuven Ramaty High Energy Solar Spectroscopic Imager (\rhessi) during the first half of 2002. For each, we examined dynamic spectra, time- and frequency-synthesis maps, \rhessi\ images with overlaid \ovsa\ contours, and a few representative single-frequency snapshot \ovsa\ images. We focus on the frequency dependence of \mw\ source sizes derived from the \ovsa\ images and their effect on the low-frequency \mw\ spectral slope. We succeed in categorizing 18 analyzed events into several groups. Four events demonstrate clear evidence of being dominated by \gs\ self-absorption, with an inferred brightness temperature of $\geq10^8$~K. 
The low-frequency spectra in the remaining events are affected to varying degree by Razin suppression. We find that many radio sources are rather large at low frequencies, which can have important implications for solar energetic particle production and escape.

\end{abstract}

\section{Introduction}

Solar flares produce a variety of transient emissions over the electromagnetic spectrum from radio waves to hard X-rays (HXR) and gamma rays.  Nonthermal electrons accelerated in the flares dominate the two ends of the spectrum, microwaves (MW) and HXRs. A typical \gs\ \mw\ spectrum has a spectral peak frequency at about $\sim10$~GHz flanked by low- and high-frequency slopes that often can be well approximated by power-laws \citep{Nita_etal_2004}.

Traditionally, the high-frequency, optically thin part of the \gs\ spectrum has received the greatest attention. Historically, the interest in this high-frequency region developed from the study of supernova remnants and radio galaxies \citep[see, e.g., ][]{1965ARA&A...3..297G}, where the power-law index $\alpha$ of the radio spectrum can be unambiguously related to the power-law index $\delta$ of the energy distribution of ultrarelativistic electrons producing synchrotron emission, $\delta=2\alpha-1$. Accordingly, a similar interpretation was attempted in early studies of the \mw\ emission from solar flares. However, it was long ago recognized \citep[e.g.,][]{Dulk_1985} that in the solar case the \mwe\ is produced by non- or only mildly relativistic electrons, for which that simple relationship between the spectral indices no longer holds. Moreover, the \gs\ spectrum from power-law distributed nonthermal electrons does not follow a simple power-law \citep{Ramaty_1969} but rather the slope decreases with increasing frequency. In addition to the dependence on the energy spectrum, the \gs\ spectral slope varies strongly in the presence of any pitch-angle anisotropy of the nonthermal electrons \citep{Fl_Meln_2003b}. Therefore, the energy spectrum of radiating electrons cannot reliably be derived from the radio slope alone. Instead, either a forward fit \citep{Gary_etal_2013} or realistic 3D modeling \citep{Nita_etal_2015} is needed to extract plasma and particle diagnostics from \mw\ emission.

With 3D modeling \citep[see, e.g.,][]{Fl_etal_2016, Fl_Xu_etal_2016,2018ApJ...852...32K} however, the strong diagnostics potential of the low-frequency part of the \mw\ spectrum becomes apparent.  Such modeling helps to recover spatial distributions of nonthermal electrons and magnetic field (and, in some cases, of the thermal plasma) in addition to the energy distribution of nonthermal electrons. In particular, the modeling vividly reveals the theoretically expected link between the \mw\ emission spectral shape at the low frequencies and the source area, which may be strongly frequency dependent \citep{Bastian_etal_1998, Kuznetsov_etal_2011}. This suggests that having simultaneous information about the \mw\ spectrum and associated source area from imaging at various frequencies would allow a more accurate interpretation of the radio spectrum and the realization of its diagnostic potential.

Imaging of the low-frequency part of the \mw\ spectrum is needed for reliable diagnostics because in addition to its area dependence, a few other effects can shape it: \gs\ self-absorption \citep{Dulk_1985} by the nonthermal electrons themselves, the suppression of the emission via the Razin-effect \citep{victor}, and free-free absorption by the thermal plasma \citep{Bastian_etal_2007}. The latter two effects require a dense plasma, with free-free absorption requiring in addition a relatively low temperature. Although in the case of a uniform source all these cases could be distinguished by studying the \mw\ spectral shape alone, the commonly occurring source non-uniformity complicates the interpretation greatly; thus, considering the spectral and spatial data together is needed to disentangle the different factors forming the \mw\ spectrum at the low frequencies.  The paucity of such spectral imaging explains the relative lack of attention that has been given to the spectrum below the peak frequency.

Surprisingly, there have been only a few isolated studies of \gs\ source sizes or locations relative to other flare-associated sources at frequencies $f\lesssim5$~GHz. \citet{Kucera_etal_1993, 1993ApJ...413..798W} reported observations with Very Large Array (VLA) at 327~MHz \& 1.4~GHz, which yielded a conclusion that there can be multiple \mw\ sources at low frequencies, with much larger total source area than is typical at higher frequencies. In their review paper, \citet{Bastian_etal_1998} used a combination of a few observations and simple models to conclude that ``typically, the characteristic source scale declines roughly as'' $f^{-1}$; however, this should only be considered a general qualitative trend, from which individual cases can deviate substantially. There has been little progress on these ideas since then, likely because of the many studies based on the solar-dedicated Nobeyama Radioheliograph (NoRH) \citep{Nakajima_etal_1994}, which operates at high, optically-thin frequencies (17 \& 34~GHz) and drew attention to diagnostics at those frequencies. But even at the frequencies of NoRH or at the lower frequency (5.7~GHz) of the Siberian Solar Radio Telescope \citep[SSRT][]{Grechnev_etal_2003SSRT}, we are aware of no statistical study on source area even though the data needed for that are available.

Another solar-dedicated instrument, Owens Valley Solar Array \citep[\ovsa,][]{Gary_Hurford_1994} provided interferometry data suitable for imaging since 2000; however, until recently only a very few solar bursts have been imaged \citep[see, e.g.,][]{Lee_Gary_2000, Nita_etal_2005, Fl_etal_2011, Fl_Xu_etal_2016} 
because of the challenge of calibrating the \ovsa\ data. Recently, however, a calibrated \ovsa\ database and corresponding analysis tools have been produced \citep{2014AAS...22421845N} enabling us and others to perform more studies of the \mw\ source areas and locations at various frequencies in a broad spectral range, $\sim1-15$~GHz. In this paper we consider a subset of solar bursts observed simultaneously with \ovsa\ and \textit{Reuven Ramaty High Energy Solar Spectroscopic Imager} \citep[\rhessi,][]{lin2002} during the first four months of the \rhessi\ operation. Our subset includes 1 M-class and 15 C-class flares; we specifically did not include any strong flares, which typically involve extreme spatial complexity. For all of these events we obtain the dependence of source area on frequency and classify the events into various groups according to which physical effects are responsible for forming the low-frequency part of the spectrum. We give a few examples of these distinctly behaving events and discuss implications of our finding for understanding solar flares.

%
%
%
%

\section{\ovsa\ Instrument: Microwave Spectroscopy \& Imaging}

At the time of observations, the \ovsa\ interferometer consisted of 6 antennas of different diameters: two large antennas with a diameter of 27 m and four small ones with a diameter of 1.8 m. Observations were carried out at 40 characteristic frequencies in the frequency range of 1--18 GHz. The study required supplementing observations in the X-ray domain to verify the correct positioning of the interferometric images. Therefore, initially the criterion for the selection of events was the presence of observational data simultaneously in \ovsa\ and \rhessi\ databases. A list of solar flares observed simultaneously by the \ovsa\ and \rhessi\ instruments was compiled, and it contains events from 2002 February 15 (the start date of \rhessi\ observations) to 2002 July 1.

Here we specifically concentrate on the microwave imaging, along with spectral radio data obtained with \ovsa.
The spectral data consist of the total power (TP) data from \ovsa\ at 40 frequencies spaced roughly logarithmically from 1--18~GHz. In particular, the peak frequency of the bursts falls within the \ovsa\ spectral range during the burst, so the spectral shape of the \mw\ emission, which includes the spectral peak along with low- and high- frequency slopes, was well captured by \ovsa\ alone.

The \ovsa\ imaging capability and the adopted calibration scheme described at the OVSA web site \citep[see also][for a brief description]{Fl_etal_2016_narrow} provides imaging of solar flares at many adjacent frequencies between $\sim$~1--15~GHz, which is a significant advance compared with other radio instruments available during solar cycle \#~23. Unfortunately, there are also disadvantages of \ovsa\ imaging, the most severe of which is a relatively small number of baselines available and, thus, greatly undersampling the $uv$ points needed for robust imaging. The relatively low sensitivity of the \ovsa\ array for calibration, combined with the complex instrumental configuration, also makes reliable imaging a significant challenge, and the quality of imaging varies from event to event.  For this reason, in this paper we rely on only the zeroth-order imaging property--that of source area--but as a check on quality we do require that source locations agree with independently observed positions from \rhessi\ as described below.

\section{Event Selection and Analysis}

\begin{figure}
    \centering
    \includegraphics[width=0.45\columnwidth, clip]{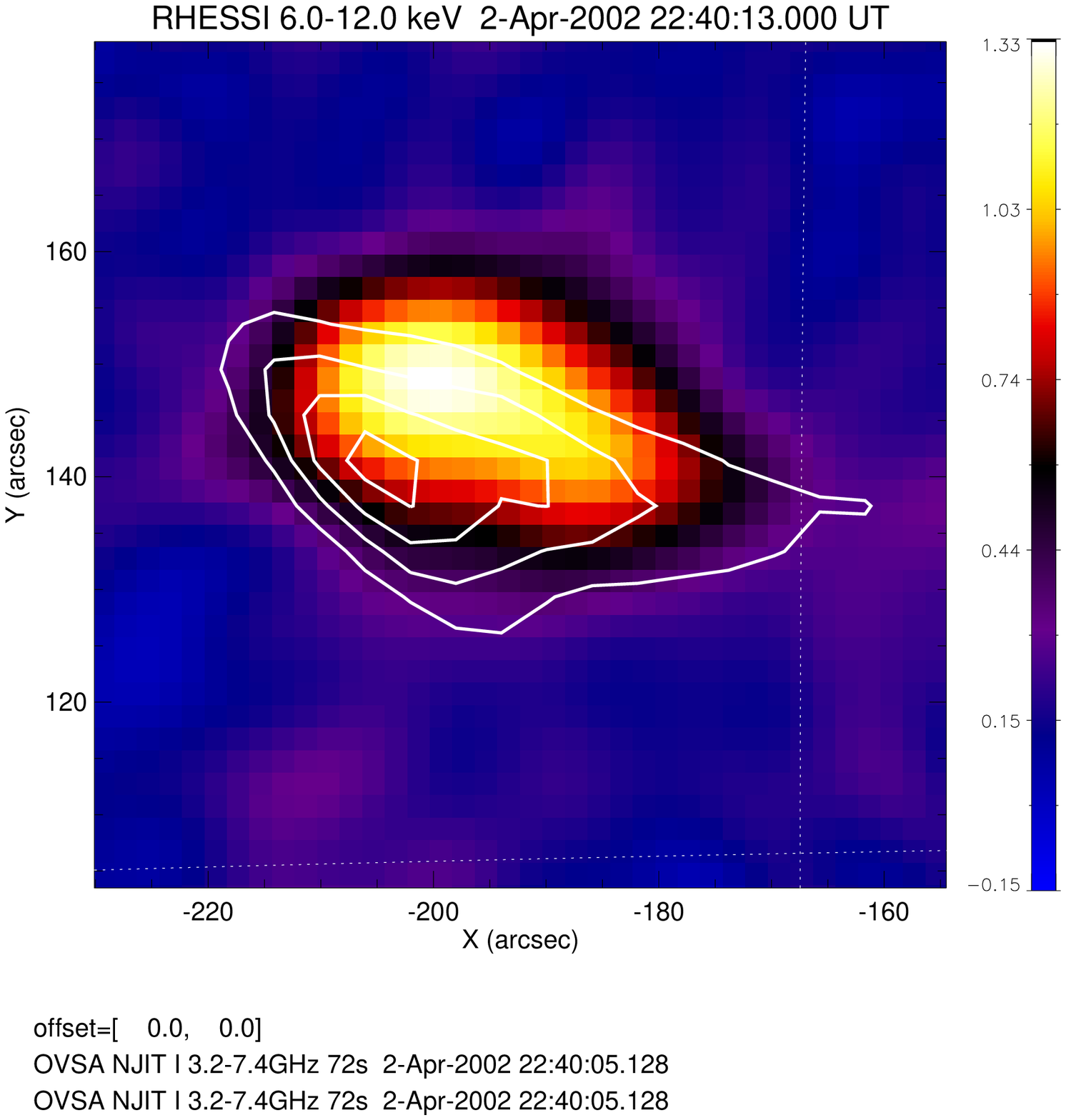}a
    \includegraphics[width=0.45\columnwidth, clip]{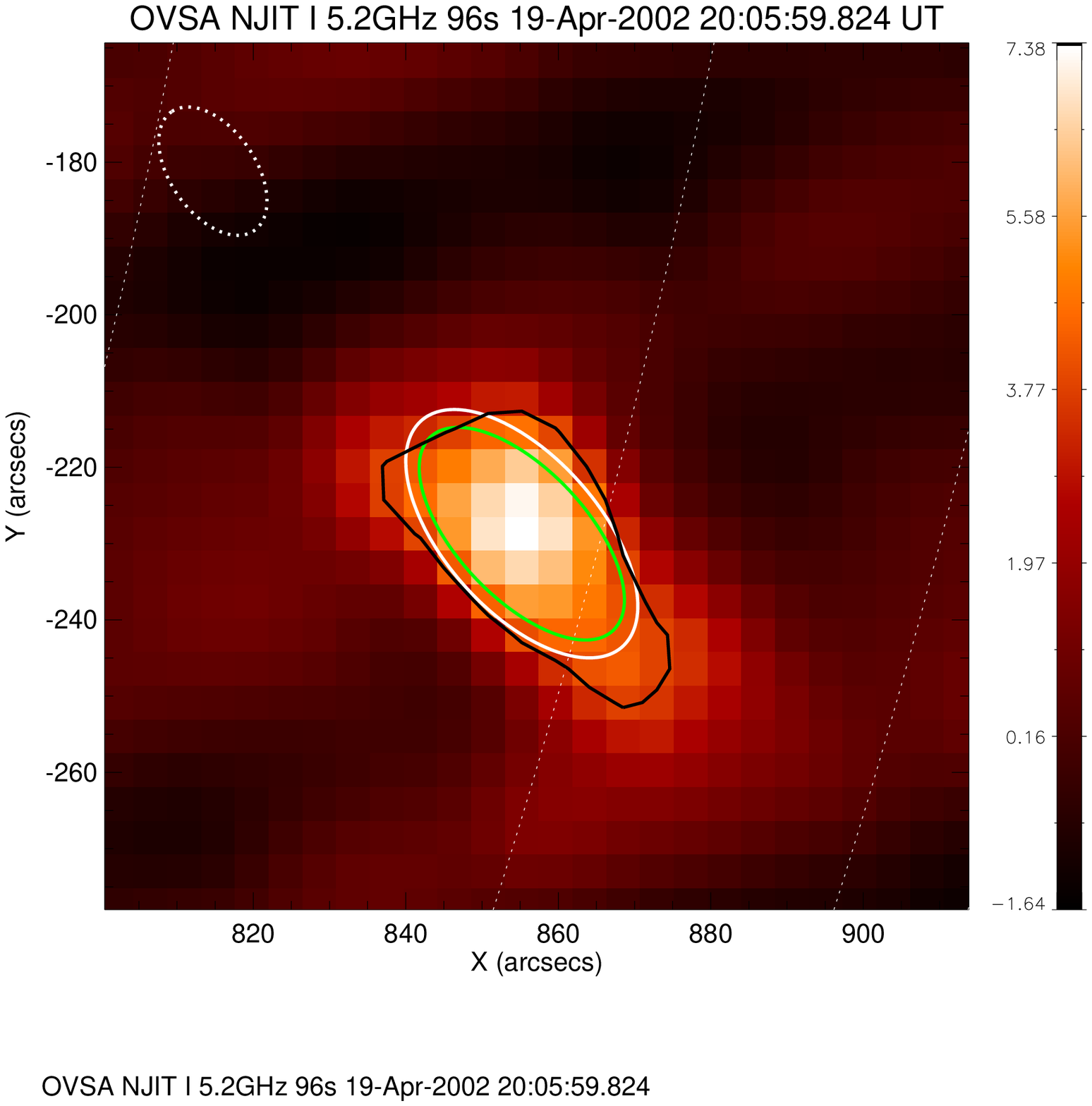}b\\
    \caption{\label{f_images} (a) An example of spatial relationship between a HXR source obtained by \rhessi\ at 6-12~keV (background image) and \mw\ OVSA image obtained by time (72~s) and frequency (3.2-7.4~GHz) synthesis using CLEAN+SelfCal method (white contours at 30\%, 50\%, 70\%, and 90\%). (b) Illustration of OVSA image and its derived quantities: 96-s-synthesis OVSA image at 5.2~GHz (red background) with 50\% intensity contour (black line); gaussian model of the OVSA beam (white dotted oval); gaussian fit to the OVSA image (white solid oval); and deconvolved gaussian source (green solid oval). }  
\end{figure}

In the period from 2002 February 15 to July 1 there are 55 events recorded by \ovsa\footnote{The initial \ovsa\ observation files in the ARC format are freely available at \url{http://ovsa.njit.edu/data/archive}.}. Of these, 37 are accompanied by observations with \rhessi. For the analysis of
the microwave source areas performed here, there is no explicit use of the \rhessi\ data aside from our use of the \rhessi\ source locations to check the validity of the OVSA imaging: the ability of the OVSA data to reproduce the correct source location (verified using direct comparison with \rhessi\ images, see Figure~\ref{f_images}) is an important initial check on the validity of the OVSA phase calibration for a given event.  From the sample of these 37 OVSA events, about half of the events are unsuitable for calibration and analysis due to incompleteness, low quality, or high noise. There remain 23 bursts for which the OVSA imaging was possible. Since in some cases there are several subbursts in the dynamic spectrum, which are treated in this study as separate events, the total number of events increases to 26. Of these 26 events, only the events with data for at least two frequencies below the peak frequency were selected in order to analyse the low-frequency part of the spectrum, which yielded the final list of 18 events listed in Table~\ref{table1}. For each of these events we apply the algorithm described in the \ovsa\ Imaging Manual\footnote{\url{http://ovsa.njit.edu/legacy/Manual4OVSA_imaging_2017.pdf}} 
to create images at as many frequencies as possible using the time synthesis technique over the duration of the event peak, and saved them in FITS format for further analysis.

\begin{figure}
    \centering
    \includegraphics[width=0.96\columnwidth, clip]{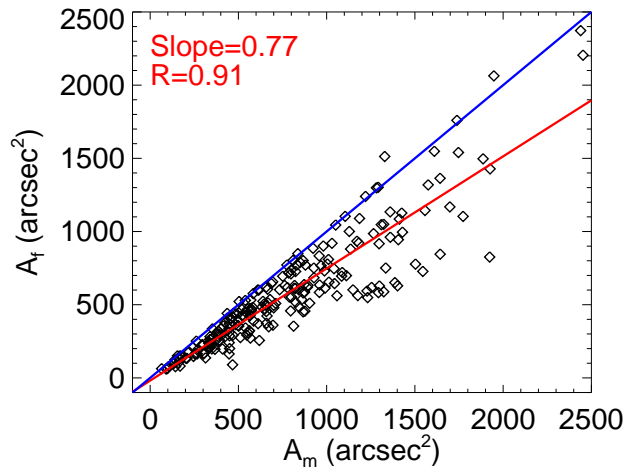}
    \caption{\label{f_scatter_areas} Scatter plot showing the areas of the gaussian fitted model vs area measured within 50\% brightness contour. The blue line shows an ideal $y=x$ correspondence, while the red line shows the cross-correlation established using the presented data, which demonstrates that the fit systematically underestimates the area, but this is typically within 25\%, according to the regression law $A_f\propto A_m$.  }  
\end{figure}

These images were used to determine the measured sources areas (biased by the frequency-dependent beam size) as well as ``deconvolved'' source areas, where the effect of the finite beam has been removed. The measured areas were straightforwardly computed by accounting for the source area within the 50\% contour.
To obtain the deconvolved areas we employed an analytical deconvolution method proposed by \citet{1970AuJPh..23..113W}, which adopts that both beam and the true source are well approximated by 2D gaussian distributions; see Figure~\ref{f_images}b for the illustration. In this case, the observed source brightness is another Gaussian, from which the true source gaussian can be determined exactly given that the beam shape is known. In our study we fit the brightness distributions for all events and at all frequencies of interest with the corresponding gaussians, and then deconvolved the beam to determine the `true' source whose area could straightforwardly be found as the ellipse area $\pi a b$, where $a$ and $b$ are the semi-major and semi-minor axes of the ellipse. Even though the gaussian fit to the source brightness distribution is not perfect in many cases, as illustrated by Figure~\ref{f_images}b, the source area is well reproduced as illustrated by Figure~\ref{f_scatter_areas}: the regression between the area of the fitted gaussian ($S_f$) and the measured (within the 50\% contour) area ($S_m$), $S_f=0.77S_m$ is highly statistically significant (the correlation coefficient is $R=0.91$); this tell us that the Gaussian fit systematically underestimates the source area by 25\% on average, which is acceptable for our study.

\begin{table*}
\caption{Events} \label{table1}      
\centering                          
\begin{tabular}{|c|c|c|c|c|c|c|c|c|}
  \hline
\#&GOES&Event&Pos&NOAA&$t_{start}$ &$t_{peak}$&$F_{peak}$,&$f_{peak}$, \\
&&&&&&&sfu &GHz\\
  \hline
1& C7.6 &	\verb"20020328_1755"&	N08W62&	9881&	17:55:31&	17:57:27&	23&	6.6\\
2& C2.3 &	\verb"20020402_2240"&	N01E12&	9887&	22:40:05&	22:40:29&	30&	5.1\\
3& C1.7 &	\verb"20020408_1842"&  N18E15&	9899&	18:43:02&	18:43:26&	27&	6.7\\
4& C2 &	\verb"20020408_1847"&	N18E15&	9899&	18:41:18&	18:47:30&	32&	6.4\\
5& C6.1 &	\verb"20020408_2038"&	N18E15&	9899&	20:40:12&	20:44:04&	20&	6\\
6& M4.0 &	\verb"20020412_1756"&	N21W25&	9901&	17:43:01&	17:56:09&	98&	4.3\\
7& M4.0 &\verb"20020412_1744"&	N21W25&	9901&	17:44:17&	17:45:18&	87&	6.2\\
8& C? &\verb"20020413_1935"&	N21W37&	9893&	19:31:48&	19:35:08&	12&	6\\
9& C2.0 &\verb"20020414_1940"&	S04E35&	9907&	19:40:27&	19:41:19&	33&	5.4\\
10& C4.4 &\verb"20020414_2054"&	S03E34&	9907&	20:50:58&	20:54:18&	21/20&	4.9/ 2.0\\
11& C4.4 &\verb"20020414_2055"&	S03E34&	9907&	20:55:06&	20:55:18&	16&	4.2\\
12& C1 &\verb"20020418_2251"&	S15W39&	9906&	22:51:17&	22:52:25&	14&	5.7\\
13& C1.6 &\verb"20020419_2004"&	S13W67&	9906&	20:04:19&	20:06:39&	24&	5.6\\
14& C3.9 &\verb"20020503_1804"&	S17E49&	9934&	18:04:47&	18:05:39&	20&	4.9\\
15& C3.2 &\verb"20020503_2333"&	S17E49&	9934&	23:32:51&	23:33:35&	345&	12.1\\
16& C2.7 &\verb"20020504_2145"&	S17E37&	9934&	21:43:05&	21:45:49&	31&	6.9\\
17& C2.4 &\verb"20020522_2050"&	S23E37&	9961& 	20:50:15&	20:51:11&	14&	4\\
18& C2.5 &\verb"20020529_1746"&	S16E62&	9973& 	17:46:24&	17:47:04&	50&	9\\
  \hline
\end{tabular}
\end{table*}

The source areas obtained this way are then fit for each event using power-laws quantified by the area extrapolated to 1~GHz and the corresponding power-law index. Although we do not have the area measurements at the reference frequency of 1~GHz, it is convenient to use this value uniformly for all events.

\section{Classification of the Events}
\subsection{Radio Brightness, Spectra, and Sizes: Theoretical Foundation}

The radio flux density $F_\sigma$ of a given polarization $\sigma$ ($= -1$ or 1 for the extraordinary [X] or ordinary [O] modes, respectively) can be computed by integrating the brightness temperature, $T_b$, of the source over the solid angle $d\Omega$ \citep{Dulk_1985, Fl_Topt_2013_CED}:

\begin{equation}
\label{Eq_flux_via T_br}
 F_{\sigma}=
 \frac{n_\sigma^2f^2}{ c^2 |\cos\theta|}k_B \int T_b d\Omega \approx $$$$ \frac{f^2}{ c^2 }k_B \int T_b  d\Omega \ \ {\rm [erg\ cm^{-2} s^{-1} Hz^{-1} ]} ,
\end{equation}
where $n_{\sigma}$ is the index of refraction of a given mode, $|\cos\theta|=\frac{c}{v_g}|\frac{\partial (\omega n_{\sigma})}{\partial \omega}|^{-1}$, $\theta$ is the angle between the wave vector and the group velocity $\mathbf{v}_g$ vector, $k_B$ is the Boltzmann constant, $c$ is the speed of light; the second (approximate) definition \citep{Dulk_1985} is valid when $n_\sigma \approx 1$ and $ |\cos\theta|\approx 1$, i.e., at  frequencies large compared with the plasma resonance frequencies.
Evaluating the integral using the Mean Value Theorem we obtain the radio flux density $F_I$  expressed in solar flux units (sfu) \citep{Fl_etal_2016}

\begin{equation}
\label{Eq_Thick_Flux}
  F_I=F_{\rm LCP}+ F_{\rm RCP}$$$$ \simeq 12~[{\rm sfu}] \left(\frac{f}{1~{\rm GHz}}\right)^{2}\left(\frac{T_{b}}{10^7~{\rm K}}\right)\left(\frac{A}{10^{20}~{\rm cm}^2}\right);
\end{equation}
where $T_{b}=T_{b}(f)$ is a peak brightness temperature of the source at the frequency $f$ and $A=A(f)$ is an effective source area at this frequency at the level of half maximum (this estimate becomes exact for a Gaussian distribution of the source brightness temperature on the sky). Note that both the brightness temperature and the source area depend on frequency in the general case.

\begin{figure}[!h]
    \centering
    \includegraphics[width=0.96\columnwidth, clip]{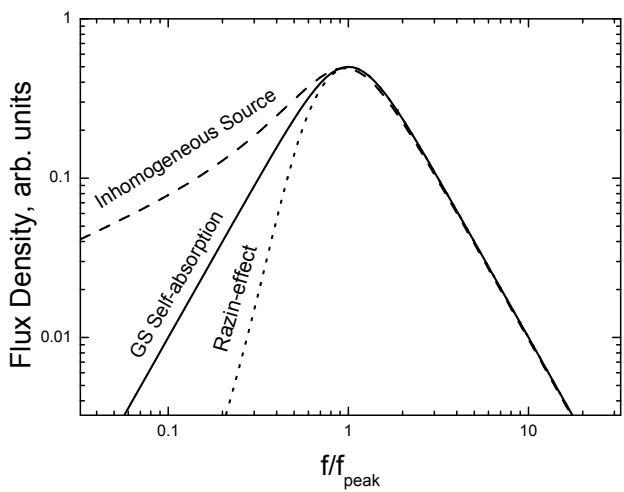}
    \caption{\label{f_cartoon} Schematic representation of the \mw\ spectral shape by Razin-effect, GS self-absorption, and source nonuniformity.  }  
\end{figure}

For a source of constant area and uniform brightness temperature, Equation~(\ref{Eq_Thick_Flux}) implies that the flux density $F_I$ increases as $f^2$.  However, actual flare spectra deviate from this expectation due to several effects shown schematically in Figure~\ref{f_cartoon}.  The largest effect is varying optical depth, which starts out high (is optically thick) at lower frequencies, then becomes unity near the peak of the spectrum (the turn-over frequency), and then falls below unity (becomes optically thin) at still higher frequencies.  In the optically thick regime, $T_b\approx T_{eff}$ \citep{Dulk_1985}, where $T_{eff}$ is the effective temperature defined by the source function $S_\sigma$: $k_B T_{eff}\approx c^2 S_\sigma/f^2$. For example, for the plasma described by a Maxwellian distribution, $T_{eff}$ is equivalent to the plasma temperature $T$. For the \gs\ emission produced by nonthermal electrons, however, the effective temperature is much higher than the plasma temperature. Given that the plasma temperature is high during flares, usually well above 10~MK, we expect still higher nonthermal brightness temperatures $T_{eff} \gg 10^7$~K.  Typically, $T_{eff} \sim 10^8-10^9$~K in solar radio bursts, although it may be even higher in strong events.  When the optical depth $\tau$ falls below unity (in the optically thin part of the spectrum), however, the brightness temperature becomes $T_b\approx \tau T_{eff}$. Other effects that cause deviations from a simple $f^2$ spectrum are that (1) $T_{eff}$ is a function of frequency for a non-Maxwellian distribution of emitting particles, generally rising with frequency due to the fact that higher-energy particles are more effective at radiating at higher frequencies,  (2) Razin-effect can modify (suppress) the low-frequency part of the spectrum if the plasma is reasonably dense, (3) the free-free absorption may play a role if the plasma is both dense and cool, and (4) due to source inhomogeneity the area of the source is a function of frequency, as already mentioned.

To distinguish between the cases dominated by either self-absorption or Razin-effect, we consider frequency dependences of all factors entering Eq.~(\ref{Eq_Thick_Flux}). Recall that, in the optically thick regime (i.e., when the self-absorption dominates), the brightness temperature follows a power-law $T_b\propto f^{\beta\lf}$ with the index $\beta\lf \sim 0.5-0.9$ for a power-law distribution of nonthermal electrons, or  $\beta\lf = 0$ for the thermal case \citep{Dulk_1985}. Thus, in a uniform source (i.e., the area does not depend on frequency), the flux density will follow a power-law $F_I\propto f^{\alpha\lf}$ with $\alpha\lf=\beta\lf+2 \sim 2.5-2.9$. The spectrum can be flatter (see illustration in Fig.~\ref{f_cartoon}) if the area increases towards lower frequencies. In contrast, within the Razin-effect, the brightness temperature falls quickly towards lower frequencies, thus, much larger values of the spectral indices, $\beta\lf>0.9$ and $\alpha\lf > 2.9$, are expected (Fig.~\ref{f_cartoon}). However, if the source is nonuniform, so that the source area falls with frequency as $A \propto f^{\gamma}$ ($\gamma$ is negative) the observed spectral index will become $\alpha\lf=\beta\lf+2+\gamma$ and, thus, can become reasonably small even under the Razin-effect.  Here the observational information about the frequency dependence of the source area can be employed to solve for a ``corrected'' spectral index $\alpha\lf_{corr}$ of the source as it would be observed if it had the same area at all frequencies, $\alpha\lf_{corr}=\beta\lf+2=\alpha\lf-\gamma$ and make classification based on this, ``corrected'' spectral index. 

Equation~(\ref{Eq_Thick_Flux}) can be solved for the brightness temperature

\begin{equation}
\label{Eq_T_br_via_Area}
  T_{b}\approx 1.56\times10^{10} [{\rm K}] \frac{F_I~[{\rm sfu}]}{f^2_{\rm [GHz]} A} 
\end{equation}
where the area $A$ is measured in arcsec$^2$, or alternatively, for the area:
\begin{equation}
\label{Eq_Area_expect}
  A\approx
  156 \frac{F_I~[{\rm sfu}]}{f^2_{\rm [GHz]} }\left(\frac{10^8~{\rm K}}{T_b}\right)
\end{equation}

Having just two measurements at a given frequency---the flux density and the source area---allows the brightness temperature to be computed using Eqn~(\ref{Eq_T_br_via_Area}).  If these are known at many frequencies, the obtained values allow the brightness temperature spectrum to be deduced, and based on that   each frequency range can be  roughly categorized as optically thick or thin.  Note that the brightness temperature of the optically thick emission is equivalent to the effective temperature (energy) of the electrons producing this emission. In solar flares, the energy of nonthermal electrons is typically above 10~keV, which implies that the expected brightness temperature is above 10$^8$~K; we will use this value as a demarcation between the optically thin and potentially optically thick regimes.

It is important to realize that it is not at all easy to reliably distinguish between the optically thin and thick cases at frequencies below the spectral peak. Indeed, the spectral turnover of the \gs\ emission can be due to either \gs\ self-absorption (the low-frequency emission is optically thick in this case) or due to suppression of the \gs\ emissivity and absorption coefficient (when the refractive index falls below unity because of high plasma density, the so-called Razin Effect; see, e.g., \citealt{Ramaty_1969}). In this case the source remains optically thin even below the spectral peak frequency and never becomes thick at the low-frequency range because the \gs\ absorption coefficient decreases almost exponentially towards lower frequencies. A combination of the two effects is also possible: the spectral peak and its immediate vicinity can be controlled by \gs\ self-absorption, while the Razin-effect can step in at lower frequencies and result in a progressively steeper low-frequency slope.

The shapes of the \gs\ spectrum in a uniform source vividly differ in the two cases  \citep[see schematic Figure~\ref{f_cartoon} and examples of the computed spectra in, e.g.,][]{Fl_Meln_2003b}. The \gs\ self-absorption results in a quasi-power-law low-frequency slope with a ``local'' spectral index slowly rising from zero (at the spectral peak) to values smaller than 3 at the lower frequencies; spectral oscillations around small integer multiples of the gyrofrequency may be present at the lowest frequencies. In contrast, the Razin-effect produces a spectrum that falls more rapidly (almost exponentially; see dotted line in Figure~\ref{f_cartoon}) towards lower frequencies. In practice, even though radio bursts with a reasonably steep low-frequency slope exist \citep{Fl_etal_2016_narrow}, in many cases, when there are independent signatures of the Razin-effect, e.g., from the spectral peak frequency constancy in time,  the spectra remain much flatter than expected for the Razin-effect \citep{victor}. The likely reason for that is well understood: it is due to a spatial non-uniformity of the radio source. In terms of Eqn~(\ref{Eq_Thick_Flux}) this non-uniformity may be described as an increase of the source area towards lower frequencies, which is the only way of compensating the drop in $T_b$ towards lower frequencies. Ideally, a combination of the spectral and imaging data can help to separate the frequency dependence of the brightness temperature from that of the source area and, thus, to classify the source as being thin or thick, which is needed to interpret the diagnostics based on the \mw\ total power data.
The physically different cases in our classification are the dominance of either effect, \gs\ self-absorption or Razin-effect, or a combination of both in either (quasi-)uniform or nonuniform source; thus, six different meaningful classes of sources are logically possible plus, perhaps, a seventh class of unclassifiable cases, e.g., in the case of spatially unresolved sources.


\subsection{Categorization of the Events}
\label{S_Categorization}

Here we will categorize the events according to the properties they demonstrate, separately---in the spatial and spectral domains. If the source area (after the beam deconvolution)  does not depend on frequency, we will classify the event as being uniform and mark it with a capital \textbf{U}; otherwise, the source is spatially complex and marked with \textbf{C}. In the spectral domain, we categorize the events according to the process responsible for forming the low-frequency slope of the spectrum (to the left from the spectral peak), we will distinguish cases dominated by self-absorption (capital \textbf{S})  or Razin-effect (capital \textbf{R}). In case of some ambiguity in assigning a given property, we put the corresponding attribute in parentheses. For example, the combination of attributes \textbf{U} in the spatial domain and \textbf{S} in the spectral domain implies that in the given event we have a spatially uniform radio source, whose low-frequency slope is formed by \gs\ self-absorption, while the attribute \textbf{(S)} tell that  dominance of the self-absorption in forming the low-frequency part of the radio spectrum is more likely than the Razin-effect; attributes \textbf{S-R} tell that we have clear indications of both self-absorption and Razin-effect in the given event, while the designations \textbf{(S)-R}/\textbf{S-(R)} tell that the presence of \gs\ self-absorption/Razin-effect is ambiguous but likely. 

\begin{figure}
    \centering
    \includegraphics[width=0.96\columnwidth]{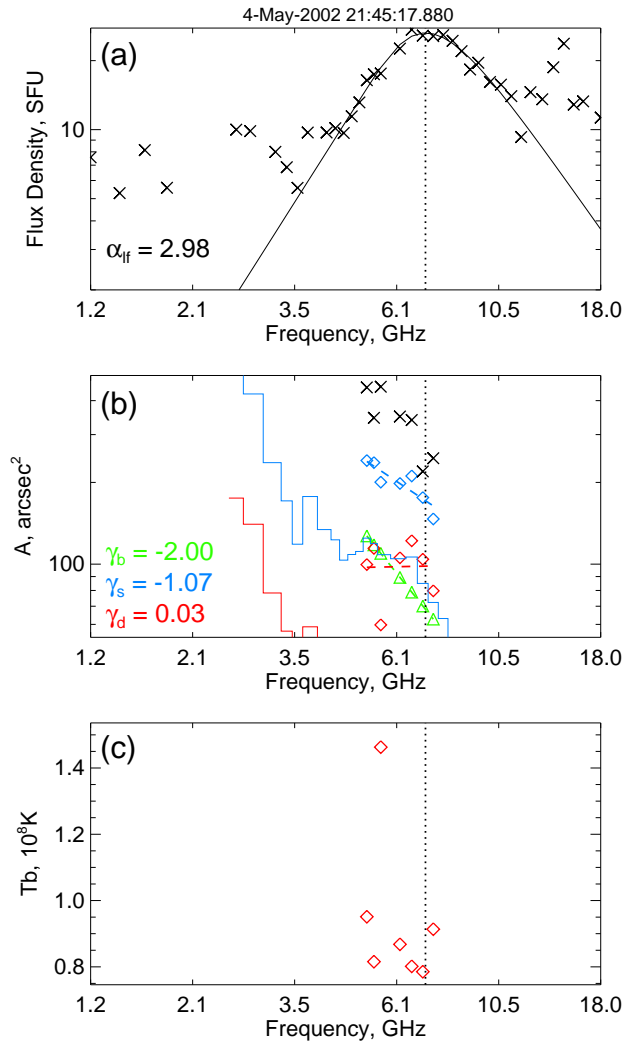}\\
    \caption{\label{f_ex_uniform} Uniform source example with a steep low-frequency slope and
almost flat dependence of the source area on frequency. Top: total power \mw\ spectrum (crosses) obtained with OVSA and spectral fit with the generic function (solid line). Middle: dependence of areas on the frequency: area measured within 50\% contour of the synthesized image (blue crosses); area of the gaussian ellipse fitted to the brightness distribution (blue diamonds); area of the OVSA array beam (green triangles); and area obtained from the gaussian deconvolved model source (red diamonds). The areas of the beam, as well as of the fitted and deconvolved sources were  computed as $\pi ab$, where $a$ and $b$ are the corresponding semi-major and semi-minor ellipse axes. Dashed blue, red, and green lines show the corresponding power-law fits to the data. Red histograms show ``expected'' area of the source computed at each frequency for the given observed radio flux using Equation~(\ref{Eq_Area_expect}) assuming that the brightness temperature at the spectral peak frequency is $T_b=10^8$~K (upper histogram) or $T_b=3\times10^8$~K (lower histogram) and the frequency dependence $T_b\propto f^{0.8}$.  Bottom: dependence of the mean brightness temperature on the frequency.
} 
\end{figure}


Table~\ref{table2} lists the results of the classification for all 18 events. It gives the spectral peak frequency, the spectral range employed for the area analysis, the low-frequency spectral indices, and the fit parameters (area at 1 GHz and the area spectral index $\gamma$) for the beam, observed source, and the deconvolved source. 
The last two columns in Table~\ref{table2} show the event attributes  according the adopted classification nomenclature.
Below we describe a number of events representative of all identified event types.


\subsection{Examples}

\begin{figure*}
\label{caseii}
    \centering
    \includegraphics[width=0.96\columnwidth]{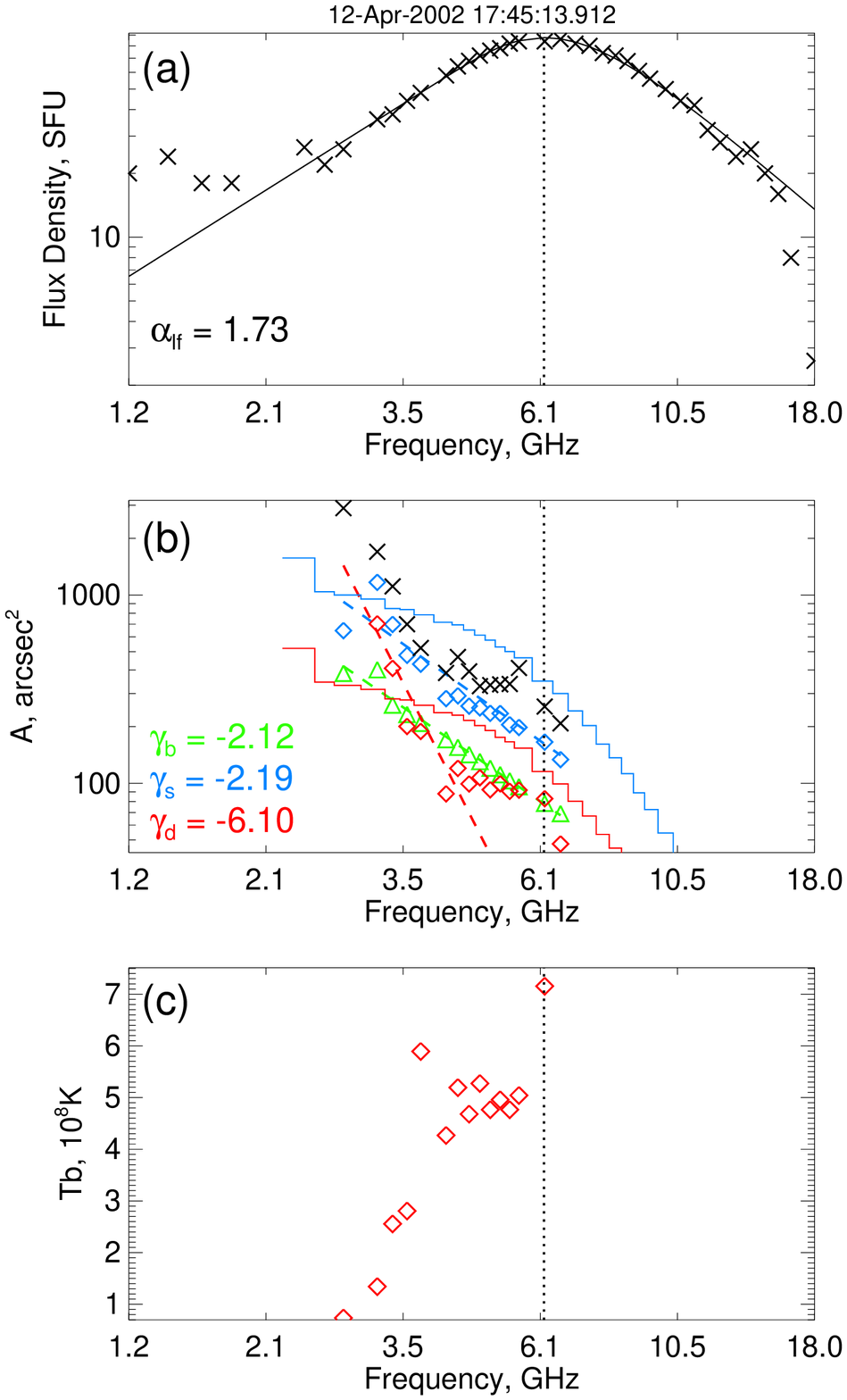}
    \includegraphics[width=0.96\columnwidth]{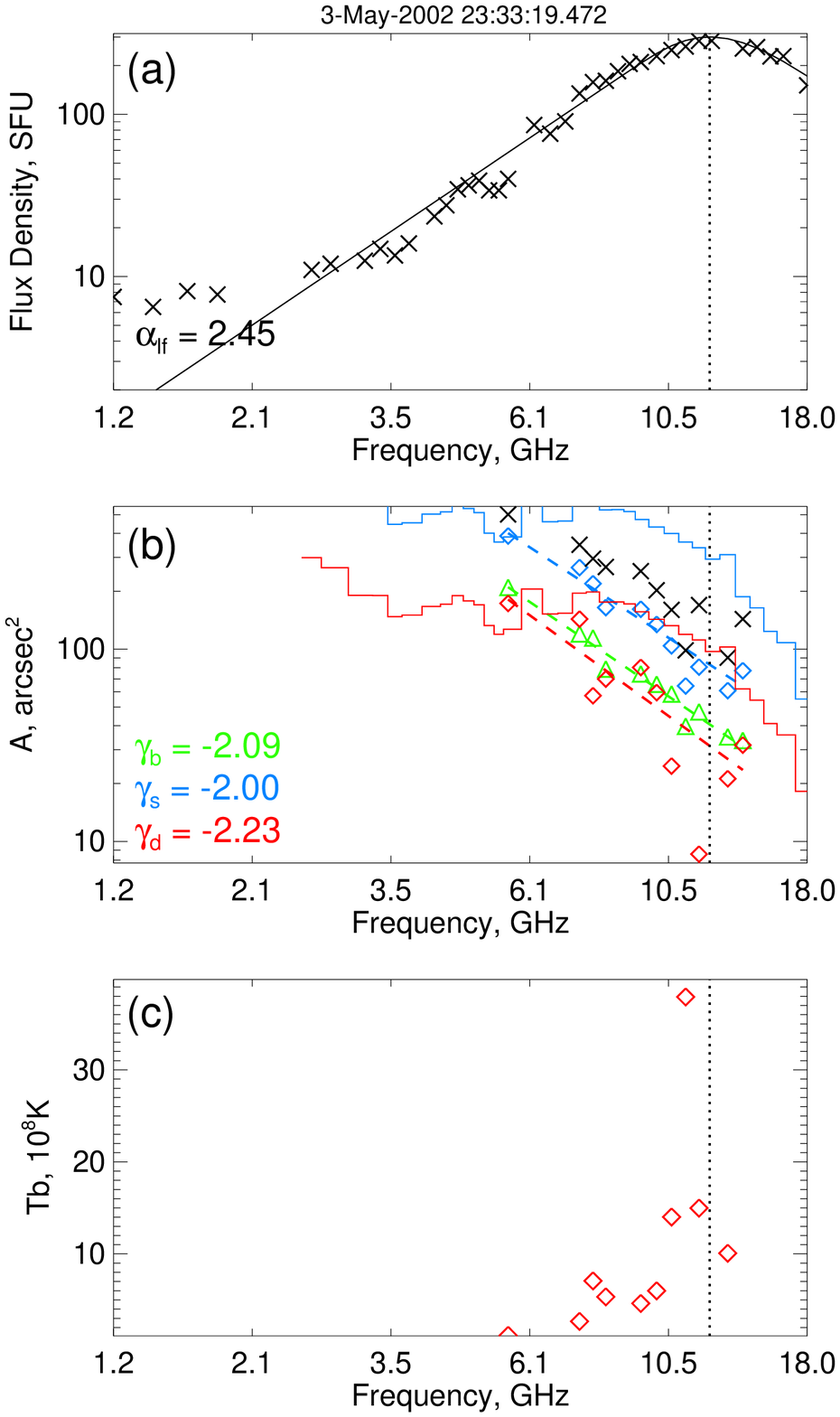}\\
    \caption{\label{f_ex_thick} Examples of optically thick, spatially nonuniform sources. The figure layout is the same as in Figure~\ref{f_ex_uniform}.
} 
\end{figure*}

\subsubsection{Example of a Uniform Source: SOL20020504T2145}

Uniform sources are, prehaps not surprisingly, very rare in our data set. Figure~\ref{f_ex_uniform} shows the spectrum and the frequency dependence of the beam and source areas for the weak \mw\ burst of solar event SOL20020504T2145. Only frequencies near the spectral peak are bright enough for reliable analysis.  The area of the deconvolved source does not show any dependence on frequency (dashed red line in Fig.~\ref{f_ex_uniform}b), which is an indication consistent with expectations for a uniform source. The low-frequency spectral slope, $\alpha\lf\approx 2.98$, is slightly steeper than the value of 2.9 expected for the optically thick spectrum; however, we are inclined to classify this case as an optically thick uniform source given that a uniform source with the Razin effect would show a progressive steepening towards the lower frequencies, which is not observed. In addition, the brightness temperature at the spectral peak frequency is about 10$^8$~K, which is consistent with the self-absorption effect. 
We have to note, however, that this consideration pertains to only the main spectral peak around 6~GHz. We cannot rule out that another spectral component is present at lower frequencies, where the signal is too low for our quantitative analysis.

\subsubsection{Nonuniform Optically Thick Sources}

Examples of the sources, where the effect of the \gs\ self-absorption is combined with nonuniformity of the source are given in Figure~\ref{f_ex_thick}; these are SOL20020412T1745 and SOL20020503T2333 events. In both cases the brightness temperature is high, $T_b\gtrsim3\times10^8$~K, while the spectral slopes are shallower than expected for a uniform source. This is consistent with the area rising towards the lower frequency as observed.

\subsubsection{Nonuniform Sources with Steep Spectra and Razin Effect Signature}

\begin{figure*}
    \centering
    \includegraphics[width=0.96\columnwidth]{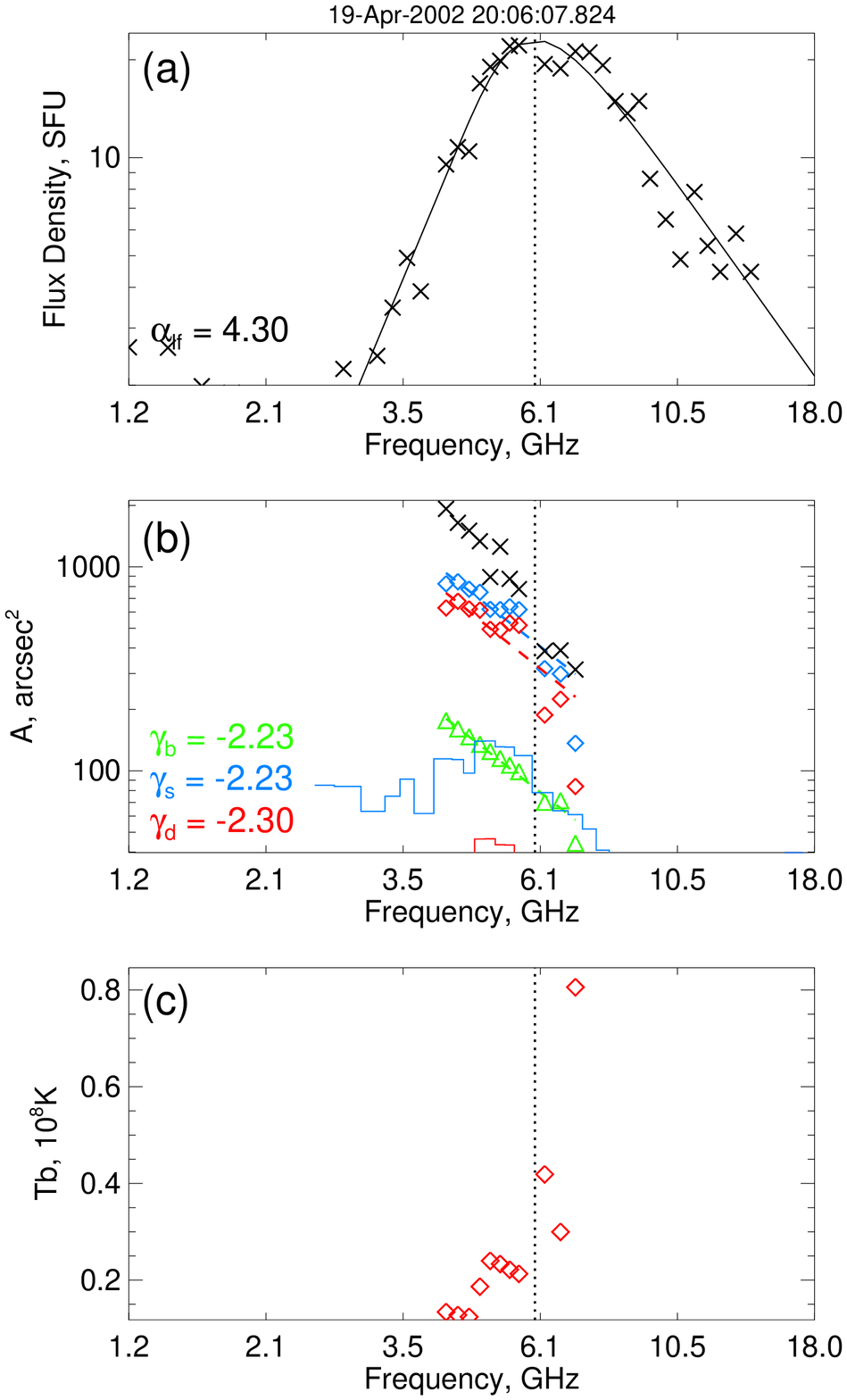}
    \includegraphics[width=0.96\columnwidth]{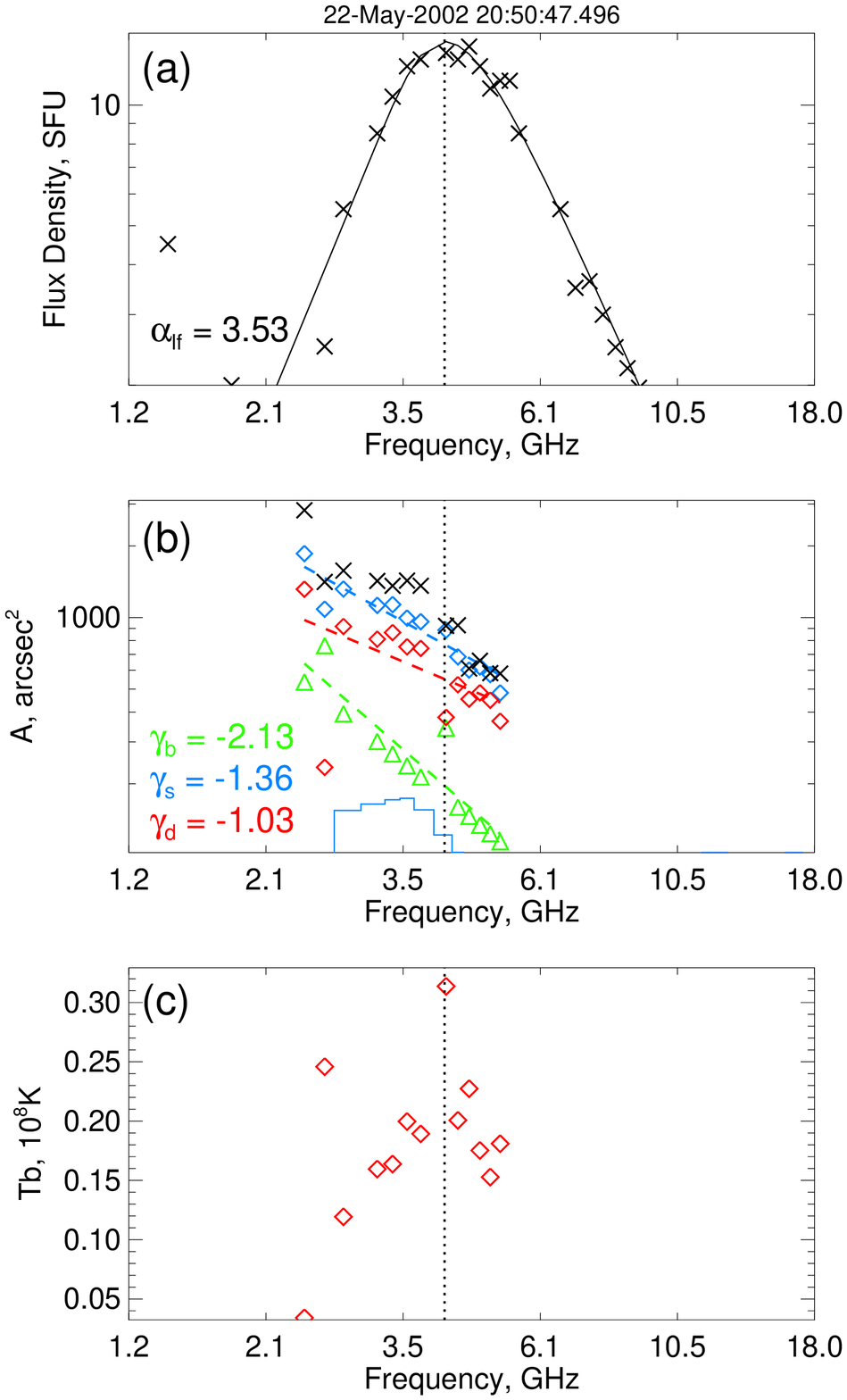}\\
    \caption{\label{f_ex_Razin_uni}
    Examples of large sources with a low brightness temperature---a signature of the Razin-effect dominance, which is also consistent with reasonably large low-frequency spectral indices of 4.3 or 3.5. The figure layout is the same as in Figure~\ref{f_ex_uniform}.
} 
\end{figure*}

Examples of sources where the Razin effect is combined with nonuniformity of the source are given in Figure~\ref{f_ex_Razin_uni};  these are the SOL20020419T2006 and SOL20020522T2050 events. In these cases the low-frequency spectral slopes are remarkably steep, $\alpha\lf = 4.3$ and 3.53, respectively; however, the deconvolved source areas still grow towards lower frequencies, which is an indication of source nonuniformity. This is, though, not surprising, given that even events with reasonably steep low-frequency spectra were found to show a significant spatial complexity \citep{Fl_etal_2016_narrow}. Here both spectral steepness and low brightness temperatures (large source areas) are consistent with the Razin effect playing the dominant role at the low frequencies.

\subsubsection{Nonuniform Sources with Shallow Spectra and Razin Effect Signature}

\begin{figure}
    \centering
    \includegraphics[width=0.96\columnwidth]{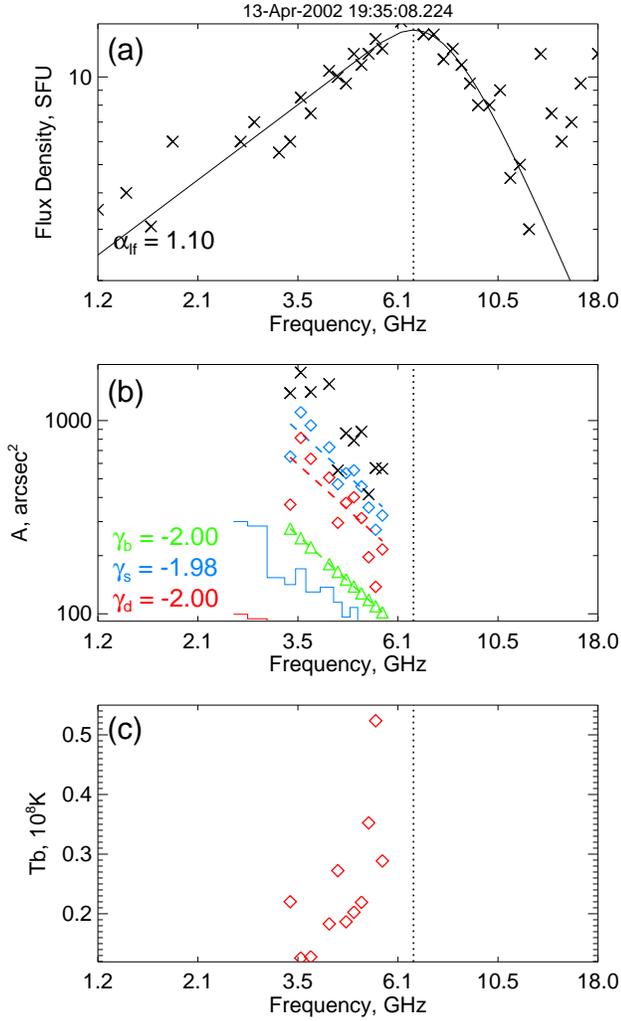}\\
    \caption{\label{f_ex_Razin_nonuni} An example of a similarly large source with a low brightness temperature as in Figure~\ref{f_ex_Razin_uni}---a signature of the Razin-effect dominance, but in this case with a relatively shallow low-frequency spectrum described by index 1.1. This is a clear indication of the spatial nonuniformity of the source, which is further confirmed by rapid increase of the source area towards lower frequencies. The figure layout is the same as in Figure~\ref{f_ex_uniform}.
} 
\end{figure}

Figure~\ref{f_ex_Razin_nonuni} shows an example (SOL20020413T1935) of a source with a shallow low-frequency spectrum, which might be mis-interpreted as \gs\ self-absorption, but whose brightness temperature is low ($\sim10^7$~K), which is more consistent with the Razin effect. Note that in a comprehensive study based on the entire 2001-2002 OVSA database, many events with quite shallow low-frequency spectra were, nevertheless, found to show signatures of the Razin effect based on their temporal evolution \citep{victor}.

\subsubsection{Nonuniform Sources with Varying Spectral Slope, Self-Absorption, and Razin Effect Signature}

\begin{figure}
    \centering
    \includegraphics[width=0.94\columnwidth]{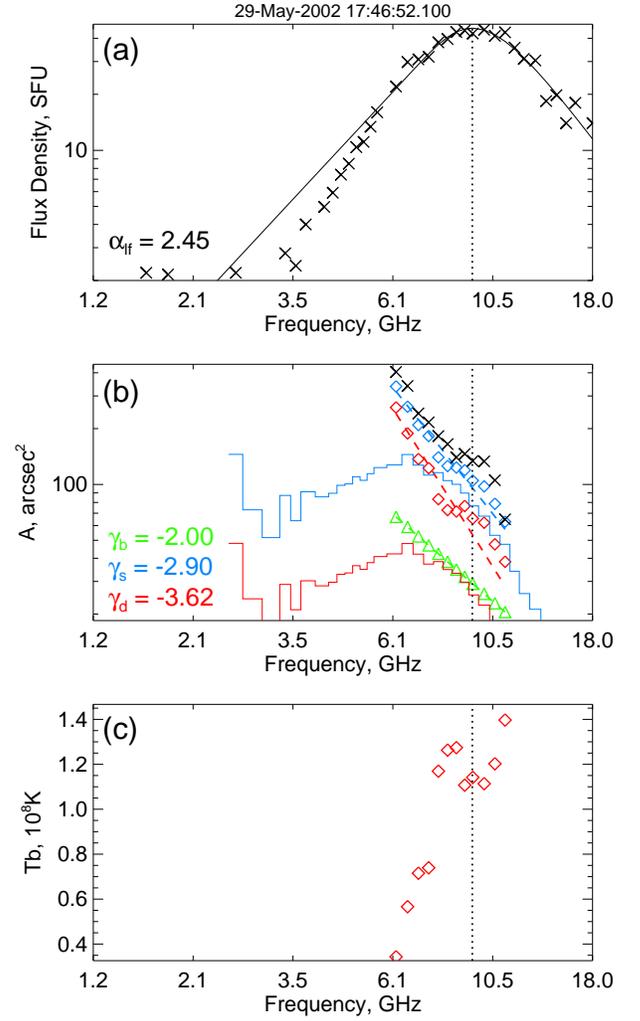}\\
    \caption{\label{f_ex_combo} 
An event showing a combination of all effects: self-absorption, Razin-effect, and spatial nonuniformity, which his vividly illustrated by the fact that the brightness temperature is high around the spectral peak but then drops very quickly towards lower frequencies. The figure layout is the same as in Figure~\ref{f_ex_uniform}.
} 
\end{figure}

An example of a case in which the \gs\ self-absorption and  Razin-effect are combined with nonuniformity of the source is given in Figure~\ref{f_ex_combo};  this is the SOL20020529T1746 event. In such cases, the low-frequency spectral slope shows a progressive steepening towards lower frequencies starting from $\alpha\lf = 2.45$. The deconvolved source area grows towards lower frequencies, which is an indication of source nonuniformity.  The brightness temperature drops significantly towards lower frequencies, which confirms the interpretation of a combination of all possible effects in these events: non-uniformity, \gs\ self-absorption, and the Razin-effect.

\section{Statistics}

\begin{figure}
    \centering
    \includegraphics[width=0.96\columnwidth]{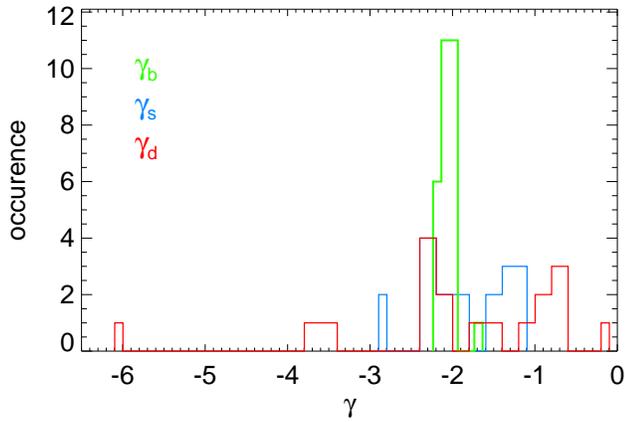}\\
\caption{\label{f_histogram_power_index} Occurrence rate of the area spectral indices $\gamma$ for the measured (blue) and deconvolved (red) sources and the OVSA beam (green) for 18 events with well presented low-frequency part of the \ovsa\ spectra. Deconvolution was performed assuming Gaussian shapes for source and beam using analytical formulae derived by \citet{1970AuJPh..23..113W}. 
} 
\end{figure}

Having the information about the spectral slope, the radio flux density, and the source areas at different frequencies, we categorized all events in Table~\ref{table1} according to the adopted classification approach, which is indicated in the last column of Table~\ref{table2}. Then, we plot the histogram of the area spectral index occurrence in Figure~\ref{f_histogram_power_index}, where we present the areas of both the Gaussian fit and the deconvolved source along with the corresponding value for the array beam. These histograms are important because it is not automatically guaranteed that OVSA imaging yields genuine source areas given the relatively small number of baselines. Nevertheless, Figure~\ref{f_histogram_power_index} shows that the frequency dependence of the beam area on the frequency follows the expected $1/f^2$ dependence, while the source areas, both gauss-fitted and deconvolved, show a significant scatter of values from 0 to $-$4, which confirms that the measured source areas are dominated by the actual source area rather than by the array beam.

\begin{figure}
    \centering
    \includegraphics[width=0.47\columnwidth]{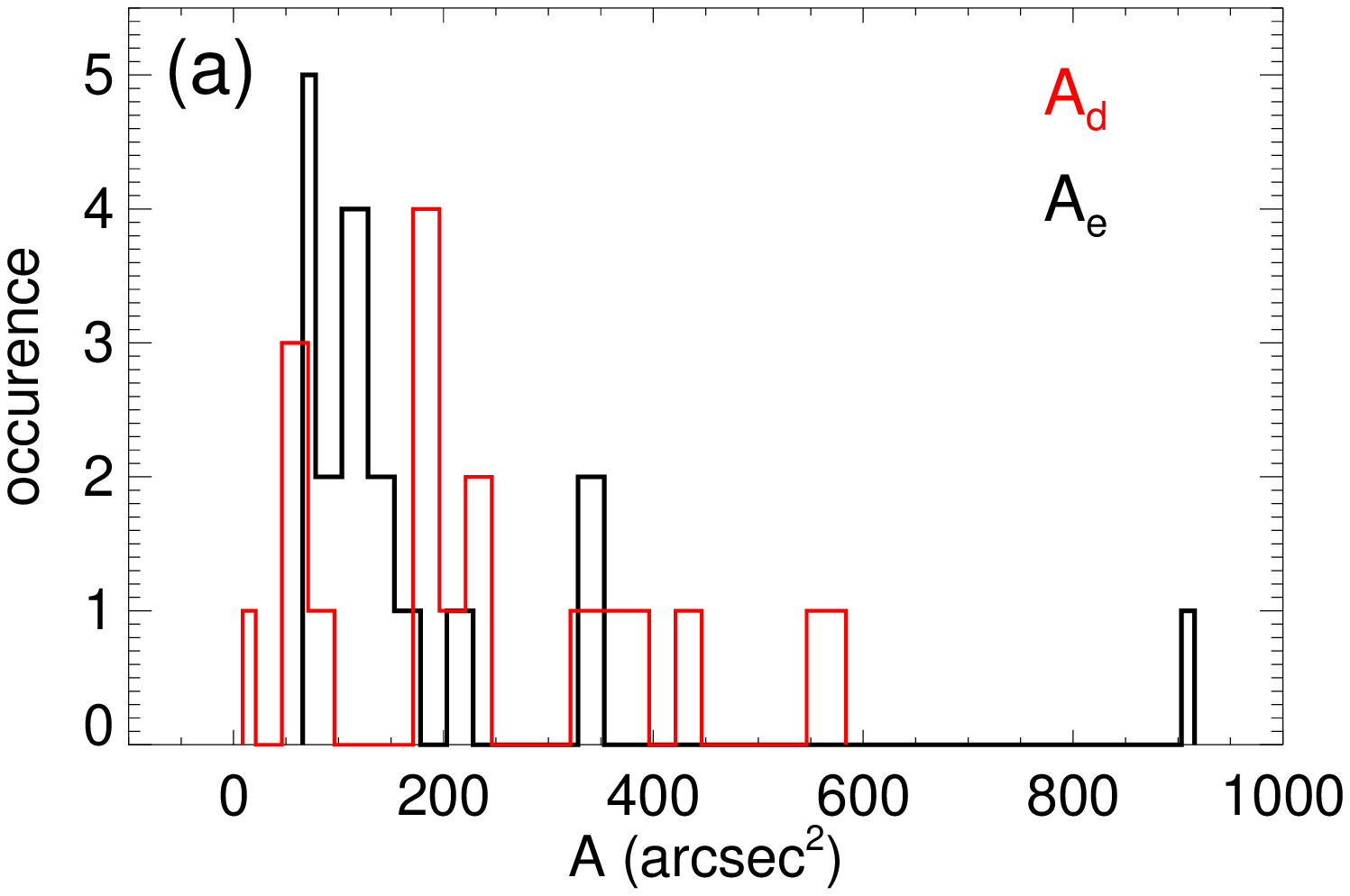}
    \includegraphics[width=0.49\columnwidth]{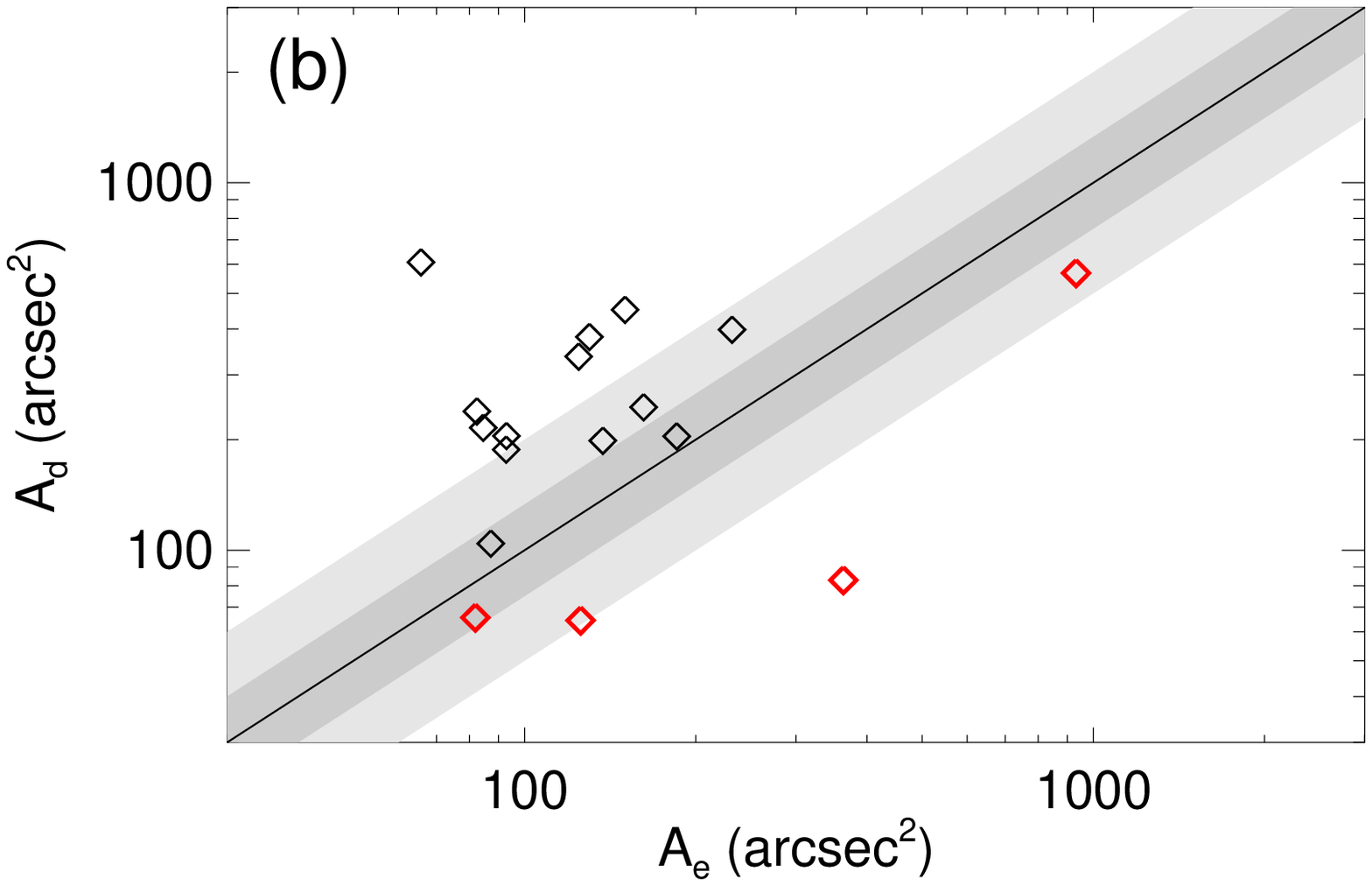}\\
    \caption{\label{f_histogram_scatter_area} Comparison of the measured and estimated source area. (a) Histograms of the deconvolved source areas (red) and estimated areas
assuming brightness
temperature $T_b = 10^8$~K (black) at the spectral peak frequency. (b)
Scatter plot of the deconvolved and estimated source areas.
Black line denotes $A_d=A_e$, while the grey shades show 25\% and 50\% deviations from that line. Events which lie below this line are plotted in
red.} 
\end{figure}

Then, Figure~\ref{f_histogram_scatter_area}a shows two histograms of the source areas at the spectral peak frequency---the expected one computed based on the radio flux density and an assumed ``typical'' brightness temperature  ($T_b=10^8$~K), and the measured one based on the imaging and image deconvolution. An interesting finding is that the measured areas after deconvolution are typically larger than the expected ones. This is further confirmed by Figure~\ref{f_histogram_scatter_area}b, where the measured areas are plotted vs the expected ones; the red dashed line shows the $y=x$ diagonal. This line can be used to split our events onto two groups: (i) the events below the line have a smaller areas than expected based on the observed flux density and reasonably modest assumption about the brightness temperature. Given that their areas are smaller than expected, we conclude that their brightness temperatures are higher than the $T_b=10^8$~K adopted to plot the diagonal line; thus, these are the optically thick cases. In contrast, (ii) the events above the diagonal line show low brightness temperature and large areas. These events are likely optically thin cases, where the spectral peak is formed due to the Razin effect. 

\begin{figure}
    \centering
    \includegraphics[width=0.96\columnwidth]{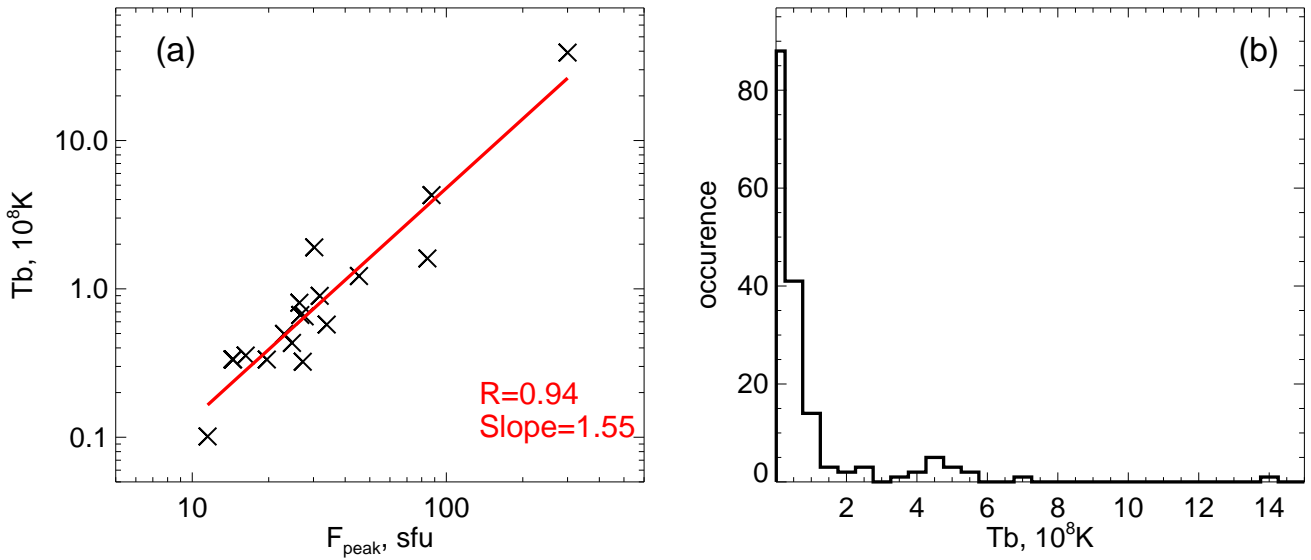}
    \caption{\label{f_histogram_Tb} (a) Scatter plot $T_b$ vs flux density at the spectral peak frequency. (b) Occurrence of the brightness temperature at all frequencies and all events considered. In the majority of cases the brightness temperature is below $10^8$~K.
    } 
\end{figure}

Figure~\ref{f_histogram_Tb}(a) shows the brightness temperature $T_b$ vs peak flux $F_{\rm peak}$ at the spectral peak frequency. It shows that these two quantities are correlated but not linearly proportional to each other: $T_b\propto F_{\rm peak}^{1.55}$ (or $F_{\rm peak}\propto T_b^{0.65}$).  This implies that the peak flux density of events increases slower than the brightness temperature, which suggests that sources with higher brightness temperature preferentially tend to have lower areas.
Figure~\ref{f_histogram_Tb}(b) shows the histogram of all brightness temperatures measured at all frequencies at and below the spectral peak. It shows that the majority of the events/frequency points show a brightness temperature below $10^8$~K, suggesting that most of them are optically thin due to the Razin effect. 

\section{Discussion \& Conclusions}

In this study we have created and made available a partial database of \ovsa\ \mw\ imaging data for 18 flares jointly observed with \rhessi. In the given study, the use of  \rhessi\ data was very limited--just to double check and validate the overall locations of the \ovsa\ images. However, the imaging data can be used in future studies employing combinations of the \mw\ and X-ray data.

This paper concentrates on such properties of the \mw\ data themselves as the frequency-dependent areas and brightness temperatures of the \mw\ sources. We found that in most cases the \mw\ sources are resolved with \ovsa\, thus, meaningful and valuable information about the source areas can be obtained from the data. Moreover, in most cases, the effect of the frequency dependent beam size of the array can be removed from the data using the analytical source deconvolution following \citet{1970AuJPh..23..113W}, which allowed us to firmly derive the brightness temperature and area separately.  Comparing with a typical brightness temperature $T_b=10^8$~K allows us to classify the sources as to whether they are suppressed by the Razin effect or not, while the frequency-dependence of the area allows us to quantify their spatial non-uniformity. Typically, the source area increases towards lower frequency as $A\propto f^{\gamma}$ with a negative $\gamma$ between $-$0.5 and $-$3.5 in agreement with a ``typical'' value of $-$2 reported by \citet{Bastian_etal_1998} but with a significant scatter. Unfortunately, we do not have sufficient statistics to firmly quantify the distribution of the  $\gamma$ indices. Most (but one) of the sources clearly show signature of spatial non-uniformity.

It is interesting to discuss the frequency dependence of the brightness temperature. As we mentioned in Section~\ref{S_Categorization}, the brightness temperature from a uniform source decreases towards lower frequency compared with the brightness temperature at the spectral peak frequency. In the data we analyzed, however, such a decrease takes place in only slightly more than half of all events (10 of 18). The rate of this decrease is observed to vary strongly from one event to the other: in some cases the slope is very shallow, while in others the brightness temperature falls rather abruptly below a certain frequency (which is an indication of the Razin effect due to dense plasma). Four events show no trend of the brightness temperature with frequency, or show a large scatter of the data points, which makes it difficult to reveal any trend. Four remaining events show the brightness temperature rising towards the lower frequencies. Having the brightness temperature rise towards lower frequencies might be explained as involving large sources of weaker magnetic field, where more energetic electrons contribute to the radio brightness \citep[e.g.,][]{2018ApJ...852...32K}.

\begin{table*}
\caption{Event classification}             \label{table2}      
\centering                          
\begin{tabular}{|c|c|cc|c|ccc|ccc|c|c|}
  \hline

\#&$f_{peak}$&\multicolumn{2}{c|}{$f_{range}$}&LF Spectral&\multicolumn{3}{c|}{Area at 1GHz}&\multicolumn{3}{c|}{Power Index}	&Case&Case\\
&GHz&GHz&GHz&Index&beam&source&dsource&beam&source&dsource&spatial&spectral\\
\hline
1&	6.6&2.6&8.6&0.97&    6727 &   25932 &   15369 &  -2.00 &  -2.18 &  -2.27&	\textbf{C} & \textbf{R}\\
2&	5.1&3.2&7.4&3.23&    4069 &    3044 &     928 &  -2.00 &  -1.21 &  -0.83&	\textbf{C} & \textbf{R}\\
3&   6.7&2.4&7.8&0.87&   4677 &    7390 &    2826 &  -2.18 &  -1.87 &  -1.62&	\textbf{C} & \textbf{S}\\
4&	6.4&3.4&7.8&0.92&    4781 &    9311 &    8227 &  -2.21 &  -1.74 &  -2.01&	\textbf{C} & \textbf{(R)}\\
5&	6&3.4&8.6&1.10 &     3195 &    3155 &     670 &  -2.00 &  -1.17 &  -0.58&	\textbf{C} & \textbf{R}\\
6&	4.3&2.8&5.6&5.44&    3728 &    8759 &  764595 &  -2.12 &  -2.19 &  -6.10&	\textbf{C} & \textbf{S}\\
7&	6.2&2.8&6.6&1.73&    2090 &    3266 &    1516 &  -1.58 &  -1.10 &  -0.87&	\textbf{C} & \textbf{S-(R)}\\
8&	6&3.4&5.4&1.10 &    3190 &   10851 &    7436 &  -2.00 &  -1.98 &  -2.00&	\textbf{C} & \textbf{R}\\
9&	5.4&3.4&7.0&2.02&    3171 &    2415 &    1223 &  -2.00 &  -0.91 &  -0.66&	\textbf{C} & \textbf{(S)-R}\\
10&	4.9/ 2.0&3.8&7.4&1.58 &    3206 &    3835 &    2247 &  -2.00 &  -1.38 &  -1.34&\textbf{C} & \textbf{R}\\
11&	4.2&2.6&4.8&2.70&    3205 &    2746 &     815 &  -2.00 &  -1.06 &  -0.56&	\textbf{C} & \textbf{R}\\
12&	5.7&3.8&6.2&2.30&    5679 &   87502 &  144338 &  -2.00 &  -2.76 &  -3.42&	\textbf{C} & \textbf{R}\\
13&	5.6&6.2&7.8&4.30&    4452 &   22904 &   20095 &  -2.23 &  -2.23 &  -2.30&	\textbf{(U)} & \textbf{R}\\
14&	4.9&3.2&7.4&3.43&    6070 &   23597 &   10799 &  -2.00 &  -2.20 &  -2.20&	\textbf{(U)} & \textbf{R}\\
15&	12.1&5.6&14.0&2.45&    7644 &   12573 &    8394 &  -2.09 &  -2.00 &  -2.2&  \textbf{C} & \textbf{S}\\
16&	6.9&4.8&7.4&2.98&    3435 &    1391 &      92 &  -2.00 &  -1.07 &   0.03&	\textbf{U} & \textbf{(S)}\\
17&	4&2.6&5.0&3.53 &    4147 &    5359 &    2409 &  -2.13 &  -1.36 &  -1.03&	\textbf{(U)} & \textbf{R}\\
18&	9&6.2&14.0&2.45&    2572 &   62510 &  179087 &  -2.00 &  -2.90 &  -3.62&	\textbf{C} & \textbf{S-R}\\
\hline
\end{tabular}
\end{table*}

Perhaps, surprisingly, only a minority (five) of the bursts show a behavior consistent with the dominance of the \gs\ self-absorption and even for some of those, a contribution from the Razin effect is likely. To some extent, this might be the outcome of our selected set of the events, which does not include particularly strong bursts, in which the self-absorption typically dominates over the Razin effect.  The majority of the bursts in our set show relatively large areas and low brightness temperatures more easily consistent with the optically thin emission due to the Razin effect. This tells us that the thermal plasma density in the \mw\ sources is relatively high in most of the flares in our sample. 

The finding that the areas of the radio sources at low frequencies (weak magnetic field) are rather large can have substantial implications for the solar energetic particle production. Indeed, these relatively large low-frequency radio sources may contain a significant number of nonthermal electrons, whose \mw\ emission is, however, weak due to the Razin effect in a source with weak magnetic field, but a reasonably dense plasma. 
We note that large low-frequency radio sources have been reported, e.g., by \citet{Fl_etal_2017,2018ApJ...852...32K}, although in their cases the radio brightness was high, likely, because of lack of the Razin-effect (due to a more tenuous plasma). This implies that having large radio sources at the low frequency can be rather common, whether the plasma density is small or large. This means that the nonthermal electrons accelerated in solar flares occupy a much larger volume that can be traced by high-frequency microwave or X-ray imaging. 

\acknowledgments

This work was supported in part by NSF grant AGS-1817277 
and NASA grants  NNX16AL67G, 
80NSSC18K0015, 
80NSSC18K0667, 
and 80NSSC18K1128
to New Jersey Institute of Technology and
RFBR grant 16-02-00749.


\bibliographystyle{apj}
\bibliography{WP_bib,solar_radio,Xray_ref,fleishman}

\end{document}